\documentclass[%
aapm,
mph,%
amsmath,amssymb,
reprint,%
]{revtex4-2}

\usepackage{graphicx}
\usepackage{dcolumn}
\usepackage{bm}

\usepackage{amsmath}
\usepackage{mathtools}
\usepackage{braket}
\usepackage{physics}
\usepackage{afterpage}
\usepackage{nicefrac}
\usepackage{xcolor}
\usepackage{amsfonts}
\usepackage{amssymb}

\usepackage[mathlines]{lineno}
\modulolinenumbers[5]
\linenumbers\relax 

\nolinenumbers
\begin{document}
	
	\preprint{AAPM/123-QED}
	
	\title{Quantum control via chirped coherent anti-Stokes Raman spectroscopy}
	
	\author{~Jabir~Chathanathil}
	\author{Svetlana~A.~Malinovskaya}%
	\affiliation{ 
		Department of Physics, Stevens Institute of Technology, Hoboken,
		NJ 07030, USA
	}%
	
	\author{Dmitry Budker}
	
	\affiliation{Johannes Gutenberg-Universit\"at Mainz, 55128 Mainz, Germany}
	\affiliation{Helmholtz-Institut Mainz, GSI Helmholtzzentrum f{\"u}r Schwerionenforschung, 55128 Mainz, Germany}
	\affiliation{Department of Physics, University of California, Berkeley, CA 94720, USA}
	
	\date{\today}
	
	\begin{abstract}
		A chirped-pulse quantum control scheme applicable to Coherent Anti-Stokes Raman Scattering  spectroscopy, named as  C-CARS, is presented aimed at maximizing the vibrational coherence in molecules. It implies  chirping of three incoming pulses in the four-wave mixing process of CARS, the pump, the Stokes and the probe, to fulfil the conditions of adiabatic passage. The scheme is derived in the framework of rotating wave approximation and adiabatic elimination of excited state manifold simplifying the four-level model system into a ``super-effective'' two level system. The robustness, spectral selectivity and adiabatic nature of this method are helpful in improving the existing methods of CARS spectroscopy for sensing, imaging and detection. We also show that the selectivity of excitation of vibrational degrees of freedom can be controlled by carefully choosing the spectral chirp rate of the pulses. 
	\end{abstract}
	
	\maketitle
	
	%
	
	Discovery of Raman scattering in the 1920s which coincided with major developments in quantum mechanics, followed by the advancements in laser technology since 1960s paved new ways for understanding the chemical compounds and structure of molecular systems. Modern Raman spectroscopy, that use spontaneous Raman scattering at their core, is characterized by a sufficiently low intensity signal, which is incoherent and isotropic. Being one of the leading non-linear optics techniques, Coherent anti-Stokes Raman Scattering (CARS) spectroscopy makes use of stimulated Raman process resulting in a directional anti-Stokes signal of intensity many orders of magnitude higher than an isotropic spontaneous Raman signal. Because CARS  addresses inherent properties of matter, such as vibrational degrees of freedom, it belongs to one of the best suited and most frequently used spectroscopic methods for imaging, sensing and detection without labeling or staining \cite{XieCRSMicroscopy, Xie_CARS_2008, Xie_CARS_2004, Quantitative_CARS_2011, Polymer_2022}. High sensitivity, high resolution and noninversiveness of CARS have been exploited for imaging of chemical and biological samples \cite{Xie_1999, Xie_imaging_2005, Xie_laser_2002, Xie_imaging_2001, Potma_hydro_2001, Potma_imaging_2006, Potma_lipid_2003, Potma_imaging_2004, Potma_imaging_2010, Saykally_imaging_2002, nature_imaging_2014}, standoff detection \cite{Dantus_standoff_2008, Dantus_standoff_2013, scully-cars, Sokolov_detection_2005} and combustion thermometry \cite{Gord_CARS_2010, Gord_comparison_2017}. Recent developments in the applications of CARS in biology include imaging and classification of cancer cells that help early diagnosis \cite{cancer1, cancer2, cancer3}, and rapid and label-free detection of the SARS-CoV-2 pathogens \cite{COVID}. CARS has also been used recently for observing real-time vibrations of chemical bonds within molecules \cite{Potma_real-time_2014}, direct imaging of molecular symmetry \cite{nature_symmetry_2016}, graphene
	imaging \cite{nature_graphere_2019}, and femtosecond spectroscopy 
	\cite{CARS_multiplex_2014, flame_wall_2015}.
	
	In this work, we present a theoretical description of a general and robust technique of creating maximal-coherence superpositions of quantum states that can be used to optimize the signals in CARS-based applications.
	
	In the four-wave mixing process of CARS, the pump and Stokes pulses having frequencies $\omega_p$ and $\omega_s$ respectively excite the molecular vibrations to create a coherent superposition which a probe pulse having frequency $\omega_{pr}$ then interacts with and generate an anti-Stokes signal. The output signal is blue-shifted and has a frequency of $\omega_{as} = \omega_p - \omega_s + \omega_{pr}$. It is about six orders of magnitude higher in amplitude compared to the spontaneous process and its direction is determined by the phase-matching condition  \cite{review_skeletal_2016, Zhetlikov_2000, Levis_BOXCARS_2007, Scully_coherent_vs_incoherent_2007}. However, one of the main challenges in CARS has been the presence of non-resonant background appearing in the spectra which limits the image contrast and chemical sensitivity. Another coherent, multi-photon Raman process called Stimulated Raman Scattering (SRS) has an advantage over CARS because of the absence of nonresonant background \cite{Xie_SRS_2008}. In SRS, the frequency difference of pump and Stokes ($\omega_p - \omega_s$) is matched with the vibrational frequency of the molecule to excite the vibrational transitions. However, the detection of compounds using SRS requires a scheme that is more complicated than CARS as the input and  output signals in SRS have the same frequencies. In CARS, the signal can be easily extracted  by an optical filter due to the anti-Stokes component having a  frequency blue-shifted compared to the incoming pulses \cite{CARS_tutorial_2012}. To overcome the limitations of CARS, there has been a tremendous effort applied on removing the background from nonresonant processes and enhancing the signal amplitude \cite{Potma_background-free_2006, Potma_background-free_2004, background-free_2003, background-free_2008, background-free_2010, background-free_2010-2, nature_high-resolution_2018, chirped-probe-pulse_2011, Pestov_Optim_2007, Kumar_background-free_2011}.
	
	In the framework of Maxwell's equations, the amplitude of the anti-Stokes field is  related to coherence between the electronic vibrational states of the target molecules. Thus, maximizing coherence is the key to optimizing the intensity of anti-Stokes signal \cite{Ch21, scully-cars}. Chirped pulses have been often used in CARS-based imaging techniques to achieve a high spectroscopic resolution \cite{Saykally_chirped_2006, Saykally_chirped_2007} and a maximum coherence \cite{Pandya20, Malinovsky_2009, Ma07}. A method for selective excitation in a  multimode system using a transform-limited pump pulse and a lineraly chirped Stokes pulse in stimulated Raman scattering is  proposed in \cite{Ma2006}. The effects of chirped pump and Stokes pulses on the nonadiabatic coupling between vibrational modes are discussed in \cite{Patel-2011}. A `roof' method of chirping to maximize coherence is introduced in \cite{Ma01} based on adiabatic passage in an effective two-level system. In this method, the Stokes pulse is  linearly chirped at the same rate as that of the pump pulse during the first half of the pulse duration and is oppositely chirped afterwards.
	
	Here, we discuss a chirping scheme in CARS, in which all the incoming pulses are chirped to achieve the maximum coherence and suppress the background via adiabatic passage. The selectivity, robustness and adiabatic nature of this control scheme make it a viable candidate for improving the current methods for imaging, sensing and detection using CARS.
	\begin{figure}
		\includegraphics[scale=0.6]{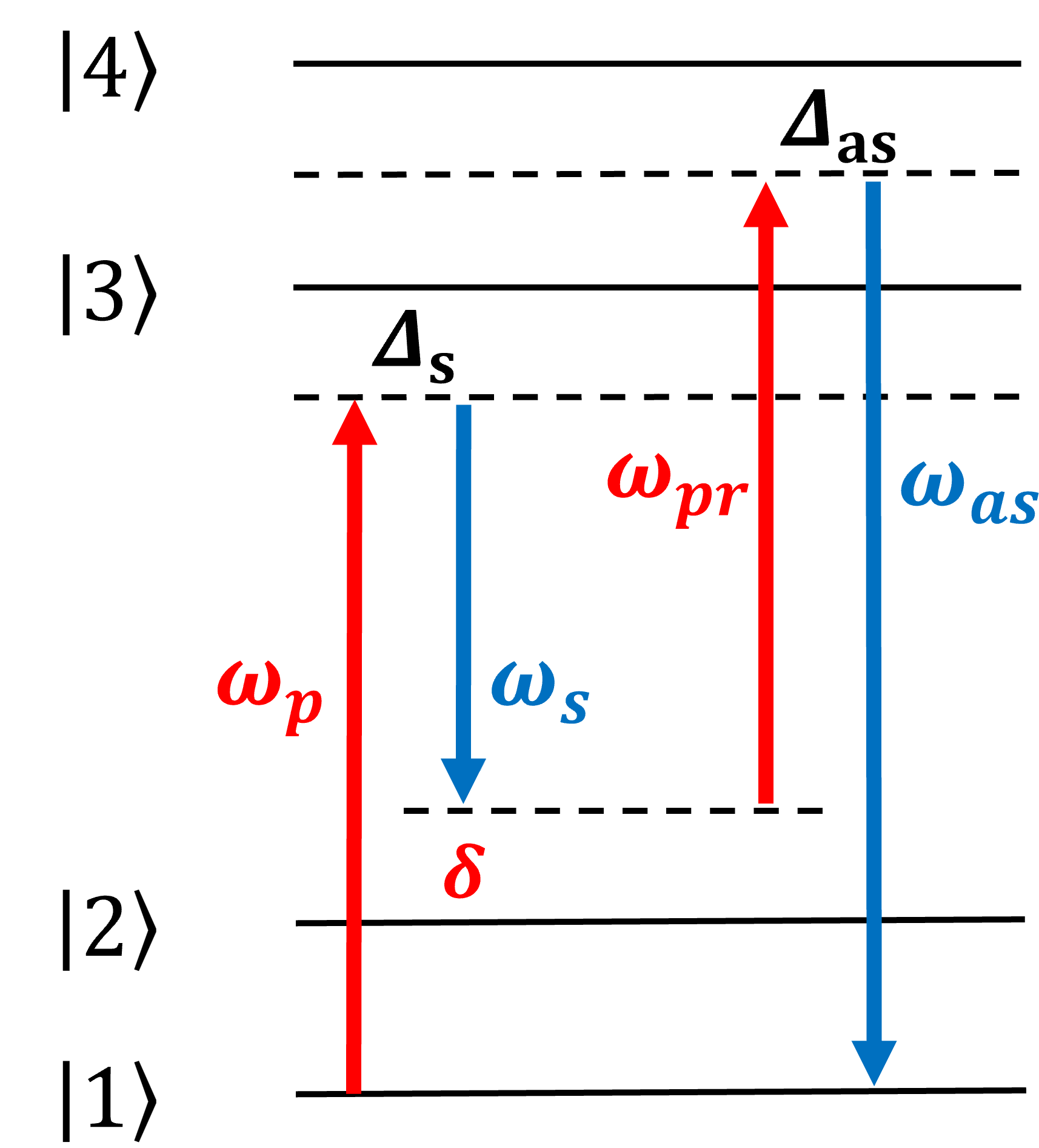}
		\caption{ Schematic of Coherent Anti-Stokes Raman Scattering (CARS), where an anti-Stokes signal is generated by the four-wave-mixing process. Maximizing the coherence between states $\ket{1}$ and $\ket{2}$, $\rho_{21}$, is the key to amplifying the signal response from the system. Here, 
			$\Delta_{s}$ and $\Delta_{as}$ are the one-photon detunings, and 
			$\delta$ is the two-photon detuning.} \label{CARS_scheme}
	\end{figure}
	
	A schematic diagram of the CARS process is given in Fig.\,\ref{CARS_scheme}. We consider chirped pump, Stokes and probe pulses with temporal chirp rates $\alpha_q$, $q=p,s,pr$ as
	\begin{equation}\label{pulses}
		E_q(t) = E_{q_0}(t)\cos\left [\omega_q(t-t_c)+\frac{\alpha_q}{2}(t-t_c)^2\right ]
	\end{equation}
	and having Gaussian envelopes
	\begin{equation}
		E_{q_0}(t) = \frac{E_{q0}}{\left (1+\frac{\alpha_q'^2}{\tau_0^4}\right )^{1/4}}e^{-\frac{(t-t_c)^2}{2\tau^2}},\label{pulses22}
	\end{equation}
	where $\tau_0$ is the transform-limited pulse duration, $\tau$ is the chirp-dependent pulse duration given by $\tau = \tau_0[1+\alpha_q'^2/\tau_0^4]^{1/2}$ and $\alpha'_q$ is the spectral chirp rate, which is related to the temporal chirp rate as $\alpha_q = \alpha_q'/\tau_0^2(1+\alpha_q'^2/\tau_0^4)$.
	The interaction Hamiltonian of the four-level system, after defining the one photon detunings, $\Delta_s = \omega_p - \omega_{31}$ and $\Delta_{as} = \omega_{as} - \omega_{41}$,
	reads
	\begin{equation}
		\begin{split}
			\mathbf{H}_{int}(t) = \frac{\hbar}{2}\left[\Omega_{p}(t) \exp(i\Delta_st - \tfrac{i}{2}\alpha_p t^2)\ketbra{1}{3} + \right.\\
			\left. \Omega_{s}(t) \exp(i\Delta_st - \tfrac{i}{2}\alpha_s t^2)\ketbra{2}{3} + \right. \\
			\left. \Omega_{pr}(t) \exp(i\Delta_{as}t - \tfrac{i}{2}\alpha_{pr} t^2)\ketbra{2}{4} + \right. \\
			\left. \Omega_{as}(t) \exp(i\Delta_{as}t)\ketbra{1}{4} + h.c \right] \,,
		\end{split}
	\end{equation}
	where $\Omega_{q0} = -\mu_{ij}E_{q0}/\hbar$. This Hamiltonian can be simplified to a two-level super-effective Hamiltonian by eliminating  states $\ket{3}$ and $\ket{4}$ adiabatically under the assumption of large one-photon detunings. The dynamics of the four-level system interacting with the fields in Eq.(\ref{pulses}) is described by the Liouville-von Neumann equation $ i\hbar \dot{\boldsymbol{\rho}}(t) = [\mathbf{H}_{int}(t), \boldsymbol{\rho}(t)].$
	We define the two-photon detuning as $\delta = \omega_p - \omega_s - \omega_{21} = \omega_{as} - \omega_{pr} - \omega_{21}$,  make the transformations in the interaction frame $\rho_{11} = \tilde{\rho}_{11}$, $\rho_{12} = \tilde{\rho}_{12}e^{-i(\omega_1-\omega_2)(t-t_c)}$, $\rho_{13} = \tilde{\rho}_{13}e^{i\omega_p(t-t_c)}$, $\rho_{14} = \tilde{\rho}_{14}e^{i\omega_{as}(t-t_c)}$, $\rho_{22} = \tilde{\rho}_{22}$, $\rho_{23} = \tilde{\rho}_{23}e^{-i(\omega_2-\omega_1-\omega_p)(t-t_c)}$, $\rho_{24} = \tilde{\rho}_{24}e^{-i(\omega_2-\omega_{1}-\omega_{as})(t-t_c)}$, $\rho_{33} = \tilde{\rho}_{33}$, $\rho_{34} = \tilde{\rho}_{34}e^{-i(\omega_p-\omega_{as})(t-t_c)}$, $\rho_{44} = \tilde{\rho}_{44}$ and obtain a system of differential equations for the density matrix elements
	\begin{align*} 
		\begin{aligned}
			i\dot{\rho}_{11} =& \tfrac{1}{2}\Omega_{p0}(t)e^{\frac{i}{2}\alpha_{p}(t-t_c)^2}\rho_{31} + \tfrac{1}{2}\Omega_{as0}(t)\rho_{41} - c.c \,, \\
			i\dot{\rho}_{22} =&    \tfrac{1}{2}\Omega_{s0}(t)e^{-i\delta(t-t_c)+\frac{i}{2}\alpha_s(t-t_c)^2}\rho_{32} \\ &+ \tfrac{1}{2}\Omega_{pr0}(t)e^{-i\delta(t-t_c)+\frac{i}{2}\alpha_{pr}(t-t_c)^2}\rho_{42} - c.c \,,\\
			i\dot{\rho}_{33} =&  \tfrac{1}{2}\Omega^*_{p0}(t)e^{-\frac{i}{2}\alpha_{p}(t-t_c)^2}\rho_{13}\\
			&+
			\tfrac{1}{2}\Omega^*_{s0}(t)e^{i\delta(t-t_c)-\frac{i}{2}\alpha_s(t-t_c)^2}\rho_{23} - c.c \,,\\
			i\dot{\rho}_{44} =&  \tfrac{1}{2}\Omega^*_{as0}(t)\rho_{14}\\ 
			&+ \tfrac{1}{2}\Omega^*_{pr0}(t)e^{i\delta(t-t_c)-\frac{i}{2}\alpha_s(t-t_c)^2}\rho_{24} - c.c \,,\\
			i\dot{\rho}_{12} =& \tfrac{1}{2}\Omega_{p0}(t)e^{\frac{i}{2}\alpha_{p}(t-t_c)^2}\rho_{32} + \tfrac{1}{2}\Omega_{as0}(t)\rho_{42} \\
			&-  \tfrac{1}{2}\Omega^*_{s0}(t)e^{i\delta(t-t_c)-\frac{i}{2}\alpha_s(t-t_c)^2}\rho_{13} \\ &-\tfrac{1}{2}\Omega^*_{pr0}(t)e^{i\delta(t-t_c)-\frac{i}{2}\alpha_{pr}(t-t_c)^2}\rho_{14} \,,\\
			i\dot{\rho}_{13} =&   \Delta_{s}\rho_{13} + \tfrac{1}{2}\Omega_{p0}(t)e^{\frac{i}{2}\alpha_{p}(t-t_c)^2}\rho_{33}\\ 
			&+ \tfrac{1}{2}\Omega_{as0}(t)\rho_{43} -  \tfrac{1}{2}\Omega_{p0}(t)e^{\frac{i}{2}\alpha_{p}(t-t_c)^2}\rho_{11} \\ &-\tfrac{1}{2}\Omega_{s0}(t)e^{-i\delta(t-t_c)+\frac{i}{2}\alpha_{pr}(t-t_c)^2}\rho_{12} \,,\\
			i\dot{\rho}_{14} =&   \Delta_{as}\rho_{14} + \tfrac{1}{2}\Omega_{p0}(t)e^{\frac{i}{2}\alpha_{p}(t-t_c)^2}\rho_{34} \\
			&+ \tfrac{1}{2}\Omega_{as0}(t)\rho_{44} -  \tfrac{1}{2}\Omega_{a0}(t)\rho_{11} \\ &-\tfrac{1}{2}\Omega_{pr0}(t)e^{-i\delta(t-t_c)+\frac{i}{2}\alpha_{pr}(t-t_c)^2}\rho_{12} \,,\\
		\end{aligned}
	\end{align*}
	\begin{align}
		\begin{aligned}
			i\dot{\rho}_{23} =& \Delta_{s}\rho_{23} +  \tfrac{1}{2}\Omega_{s0}(t)e^{-i\delta(t-t_c)+\frac{i}{2}\alpha_s(t-t_c)^2}\rho_{33} \\
			&+\tfrac{1}{2}\Omega_{pr0}(t)e^{-i\delta(t-t_c)+\frac{i}{2}\alpha_{pr}(t-t_c)^2}\rho_{43} \\ &-\tfrac{1}{2}\Omega_{p0}(t)e^{\frac{i}{2}\alpha_{p}(t-t_c)^2}\rho_{21}\\
			&-\tfrac{1}{2}\Omega_{s0}(t)e^{-i\delta(t-t_c)-\frac{i}{2}\alpha_s(t-t_c)^2}\rho_{22} \,,\\		
			i\dot{\rho}_{24} =& \Delta_{as}\rho_{24} +  \tfrac{1}{2}\Omega_{s0}(t)e^{-i\delta(t-t_c)+\frac{i}{2}\alpha_s(t-t_c)^2}\rho_{34} \\
			&+\tfrac{1}{2}\Omega_{pr0}(t)e^{-i\delta(t-t_c)+\frac{i}{2}\alpha_{pr}(t-t_c)^2}\rho_{44} \\ &-\tfrac{1}{2}\Omega_{as0}(t)\rho_{21} \\
			&-\tfrac{1}{2}\Omega_{pr0}(t)e^{-i\delta(t-t_c)+\frac{i}{2}\alpha_{pr}(t-t_c)^2}\rho_{22} \,,\\
			i\dot{\rho}_{34} =&  (\Delta_{as}-\Delta_{s})\rho_{34} + \tfrac{1}{2}\Omega^*_{p0}(t)e^{-\frac{i}{2}\alpha_{p}(t-t_c)^2}\rho_{14} \\
			&+\tfrac{1}{2}\Omega^*_{s0}(t)e^{i\delta(t-t_c)-\frac{i}{2}\alpha_s(t-t_c)^2}\rho_{24} \\ &-\tfrac{1}{2}\Omega_{as0}(t)\rho_{31} \\
			&-\tfrac{1}{2}\Omega_{pr0}(t)e^{-i\delta(t-t_c)+\frac{i}{2}\alpha_{pr}(t-t_c)^2}\rho_{32} \,.\\
		\end{aligned}
	\end{align}
	As the next step, the condition for chirping of the probe pulse such that $\alpha_{pr} = \alpha_s - \alpha_p$ is imposed,  which is required to equate the exponentials. The above set of equations is then modified by applying the adiabatic elimination, $\dot{\rho}_{33}=\dot{\rho}_{44}=\dot{\rho}_{34} = 0$ and substituting for $\rho_{14}, \rho_{24}, \rho_{23}$ and $\rho_{24}$ in the equations of $\dot{\rho}_{11}, \dot{\rho}_{22}$ and $\dot{\rho}_{12}$. After defining the new Rabi frequencies as
	
	\begin{align}
		\begin{aligned}
			\Omega_{1}(t)&=\frac{|\Omega_{p0}(t)|^{2}}{4\Delta_s}+\frac{|\Omega_{as0}(t)|^{2}}{4\Delta_{as}} \,, \\
			\Omega_{2}(t)&=\frac{|\Omega_{s0}(t)|^{2}}{4\Delta_s}+\frac{|\Omega_{pr0}(t)|^{2}}{4\Delta_{as}} \,, \\
			\Omega_{3}(t)&=\frac{\Omega_{p0}(t)\Omega^*_{s0}(t)}{4\Delta_s}+\frac{\Omega^*_{pr0}(t)\Omega_{as0}(t)}{4\Delta_{as}}
		\end{aligned}
	\end{align}
	
	the density matrix equations are cast into the a set of equations describing the dynamics in the super-effective two-level system
	\begin{align}
		\begin{aligned}
			i\dot{\rho}_{11} =& \Omega_3(t)e^{i\delta(t-t_c)-\frac{i}{2}(\alpha_s - \alpha_{p})(t-t_c)^2}\rho_{21} - c.c \,, \\
			i\dot{\rho}_{22} =& \Omega^*_3(t)e^{-i\delta(t-t_c)+\frac{i}{2}(\alpha_s - \alpha_{p})(t-t_c)^2}\rho_{12} - c.c \,, \\
			i\dot{\rho}_{12} =& \left[\Omega_1(t)-\Omega_{2}(t)\right]\rho_{12}\\ 
			&+ \Omega_3(t)e^{i\delta(t-t_c)-\frac{i}{2}(\alpha_s - \alpha_{p})(t-t_c)^2}(\rho_{22}-\rho_{11})
		\end{aligned}
	\end{align} 
	
	Further transformations of the density matrix elements such as  $\rho_{11}=\tilde{\rho}_{11}$, $\rho_{12}=\tilde{\rho}_{12} e^{i\delta(t-t_c)-\frac{i}{2}(\alpha_s - \alpha_{p})(t-t_c)^2}$, and $\rho_{22} = \tilde{\rho}_{22}$, and shifting the diagonal elements lead to the following super-effective Hamiltonian for the two-level system in the field-interaction representation
	
	\begin{widetext}
		\begin{equation}
			\small
			\mathbf{H}_{se}(t) = \frac{\hbar}{2}\left( \begin{array}{cc} \delta-(\alpha_s-\alpha_p)(t-t_{c})+\Omega_{1}(t)-\Omega_{2}(t)  & 2\Omega_{3}(t)\\ 
				2\Omega^*_{3}(t) & -\delta+(\alpha_s-\alpha_p)(t-t_{c})-\Omega_{1}(t)+\Omega_{2}(t)  \\
			\end{array} \right)\,.
			\label{Ham2level}
		\end{equation}
	\end{widetext}
	
	The amplitudes of incoming fields are manipulated to make the AC Stark shifts equal,
	$\Omega_1(t) = \Omega_2(t)$, and cancel out. This condition is  satisfied by taking $\Omega_{s0} = \Omega_{pr0} = \Omega_{p0}/\sqrt2$, considering the fact that the anti-Stokes signal is absent in the beginning of the process, $\Omega_{as0} = 0$. The effective Rabi frequency $\Omega_3(t) $  reads 
	\begin{equation}
		\Omega_3(t) = \frac{\Omega_{3(0)}}{\left[(1+\frac{\alpha_p'^2}{\tau_0^4})(1+\frac{\alpha_s'^2}{\tau_0^4})\right]^{1/4}}e^{-\frac{(t-t_c)^2}{\tau^2}}\,.
	\end{equation}
	
	\begin{figure}
		\includegraphics[scale=0.65]{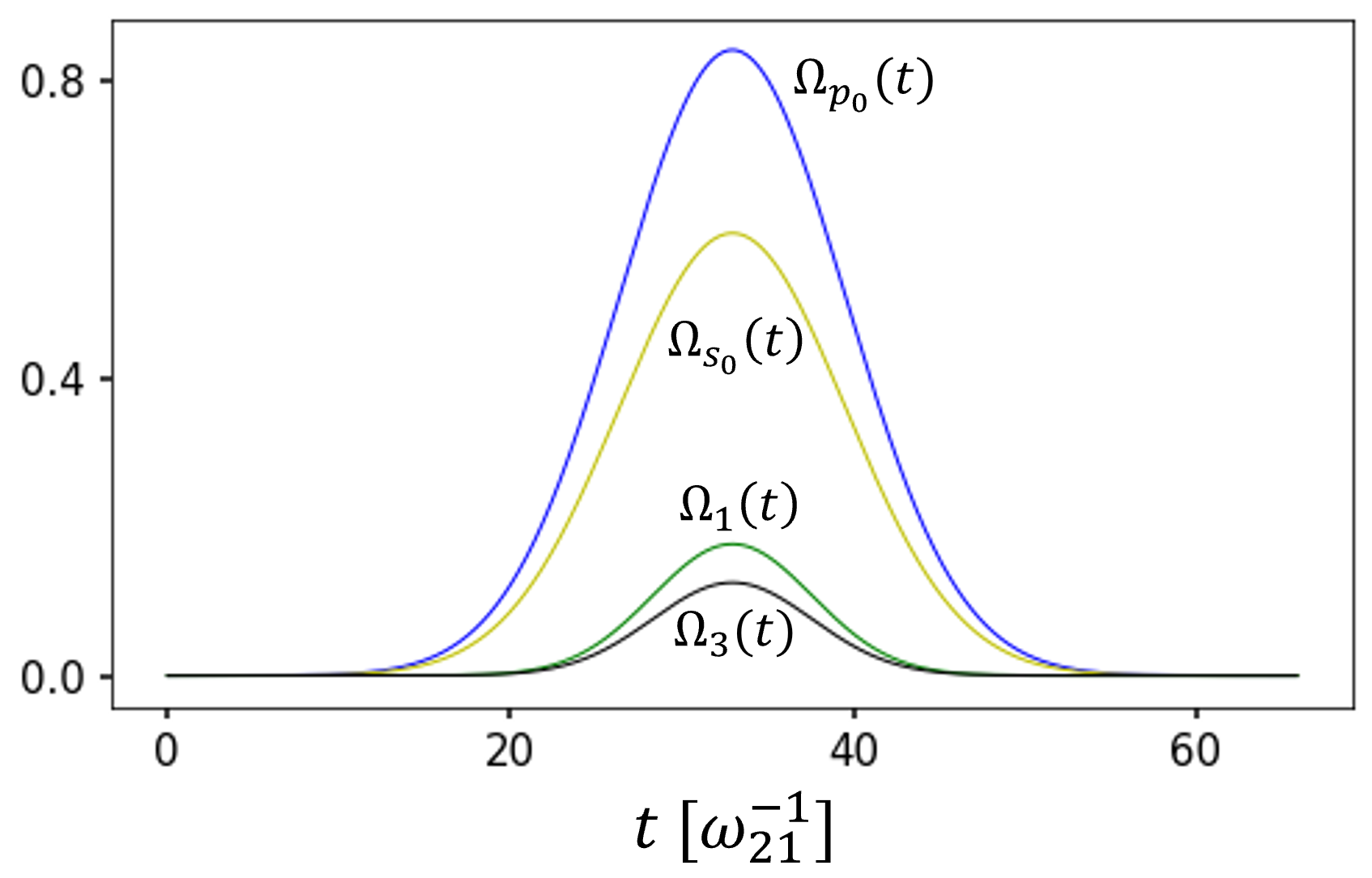}
		\caption{The evolution of different Rabi frequencies in C-CARS scheme. The Stokes and probe Rabi frequencies have the same amplitude which is less than the amplitude of the pump pulse by a factor of $\sqrt{2}$. $\Omega_{1}(t)$ and $\Omega_{2}(t)$ are canceled in the Hamiltonian making $\Omega_{3}(t)$ the only relevant quantity in the scheme.
		}\label{pulses}
	\end{figure}
	The peak effective Rabi frequency of  transform-limited pulses   $\Omega_{3(0)}$ is given by  $\Omega_{3(0)} = \Omega_{p0}^2/(4\sqrt{2}\Delta)$. It is reduced when chirping is applied to the pump and Stokes pulses with the spectral rates $\alpha_p^{\prime}$ and $\alpha_s^{\prime}$ respectively. The relative amplitudes of all the Rabi frequencies involved in the dynamics of are shown in Fig.\,\ref{pulses}. In this paper, all the frequency parameters are defined in the units of the frequency $\omega_{21}$ and time parameters are defined in the units of $\omega^{-1}_{21} $.
	
	\begin{figure*}
		\includegraphics[scale=0.85]{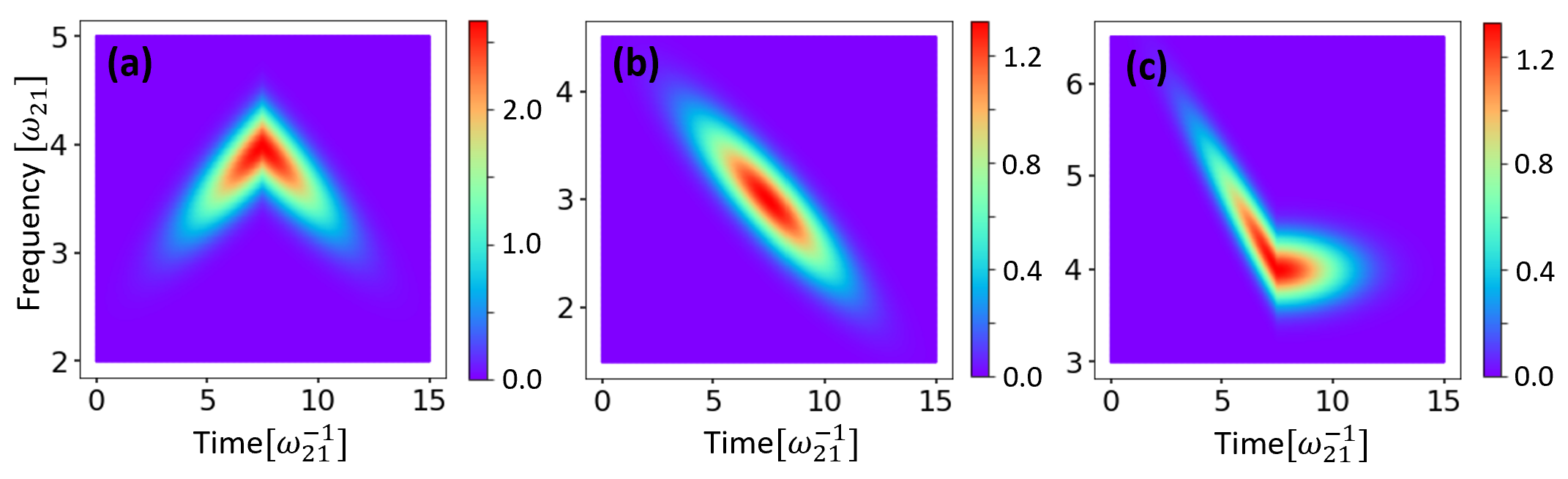}
		\caption{Wigner plots of the incident pulses; pump(a), Stokes(b) and probe(c). Note that the Stokes and probe have the same amplitudes, which is different from that of pump. The parameters used in this figure are: $\omega_p = 4.0[\omega_{21}]$, $\omega_s = 3.0[\omega_{21}]$, $\omega_{pr} = 4.0[\omega_{21}]$, $\tau = 3.0[\omega_{21}^{-1}]$, $\alpha_s = -0.2[\omega_{21}^{2}]$, and $t_c = 7.5[\omega_{21}^{-1}]$.}
		\label{wigner}
	\end{figure*}
	
	During the interaction, at time $t = t_c + \delta/(\alpha_s - \alpha_p)$, the diagonal elements become equal to zero creating a coherent superposition state having equal populations of  states $\ket{1}$ and $|\ket{2} $ and, therefore, a maximum value of coherence $\rho_{21}$. This time can be determined for a fixed value of two-photon detuning, $\delta$. At the two-photon resonance, the system reaches this maximum coherence at the central time $t_c$. It is further preserved in the state of the maximum coherence by imposing the condition of $(\alpha_s - \alpha_p)=0$  in the second half of the pulse duration. A smooth realization of this scheme is possible  by choosing the temporal chirp rates of the pump and Stokes pulses to be the same in magnitude and opposite in sign before the central time and equal in sign after that, along with the condition imposed for the chirp rate of probe pulse such that $\alpha_{pr} = \alpha_s - \alpha_p$, which was deduced to arrive into the Hamiltonian in Eq.\,\eqref{Ham2level}. 
	This chirping scheme, namely C-CARS, is summarized as follows: $\alpha_p = -\alpha_s$ and $\alpha_{pr} = 2\alpha_s$ for $t \leq t_c$, and $\alpha_p = \alpha_s$ and $\alpha_{pr} = 0$ for $t > t_c$. The Wigner-Ville distributions of the incident pulses  are found to be
	\begin{equation}
		\begin{split}
			W_{E_q}(t,\omega) = \frac{\tau\sqrt{\pi}}{2}E_{q_0}e^{-(t-t_c)^2/\tau^2}\left[e^{-\tau^2[\omega-\omega_q-\alpha_q(t-t_c)]^2} \right. \\
			\left.+ e^{-\tau^2[\omega+\omega_q+\alpha_q(t-t_c)]^2}\right].
		\end{split}
	\end{equation}
	The positive solutions of these equations are depicted in Fig.\,\ref{wigner}. The ``turning off'' of chirping in the second half of the pulse duration is the essence of this scheme, resulting in a selective excitation of the molecules and suppressing any off-resonant background. If $\alpha_p$ is not reversed, the coherence is not preserved leading to population inversion between states $\ket{1}$ and $\ket{2}$.
	
	\begin{figure}
		\includegraphics[scale=0.34]{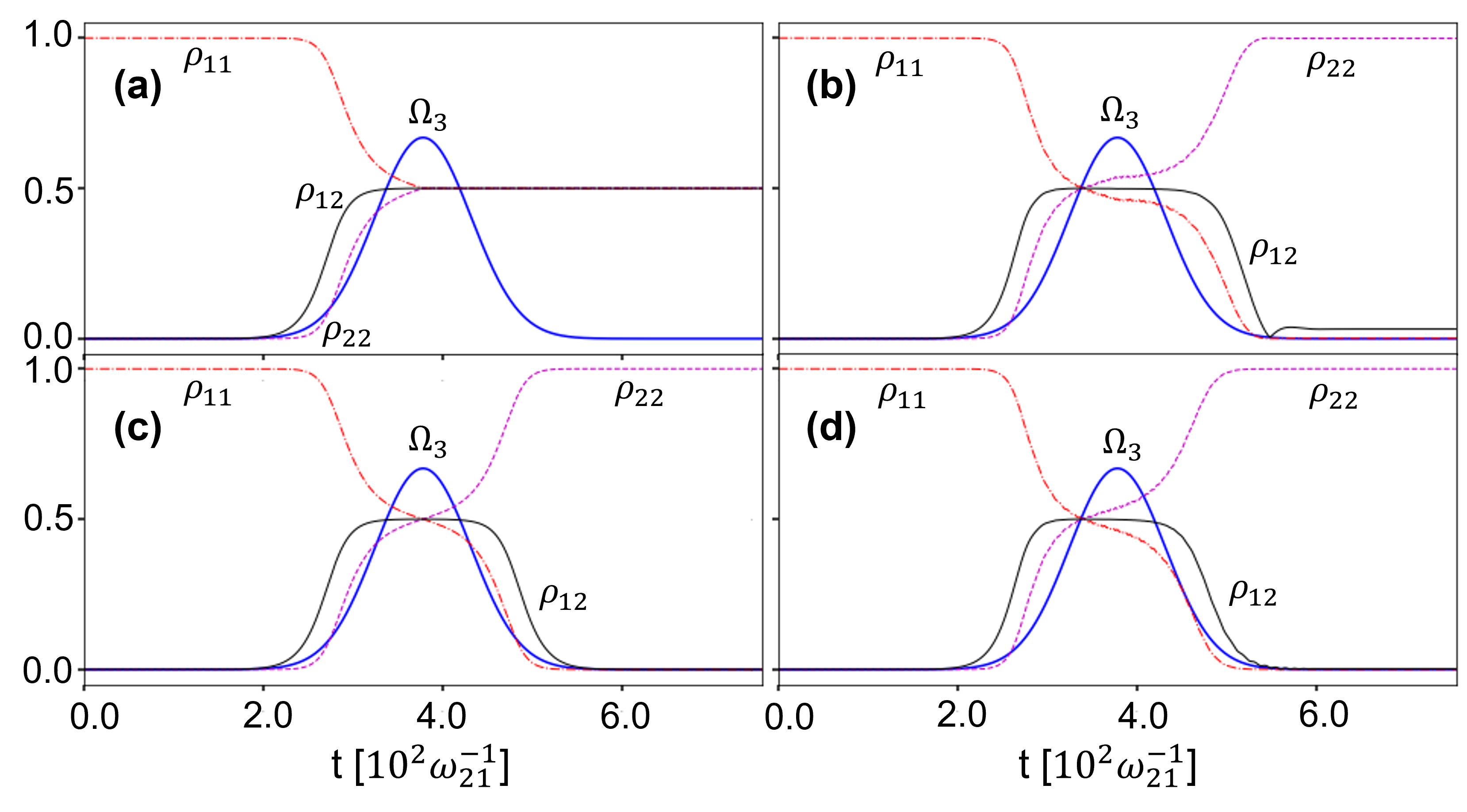}
		\caption{The evolution of the populations and coherence demonstrating selective coherent excitation in C-CARS: in (a) and (b), C-CARS scheme is applied to the resonant case ($\delta =0$)  (a)  and  off-resonant ($\delta = 0.1$) case (b). Coherence is preserved at the maximum value in the case of resonance, while it is destroyed in the detuned case. This is in contrast with the chirping scheme where the pump and Stokes pulses are oppositely chirped, $\alpha_{p} = -\alpha_{s}$, for the whole pulse duration, shown for the resonant case ($\delta =0$) in (c)  and  off-resonant ($\delta = 0.1$) case in (d). The dynamics is similar and the coherence is zero in both these cases demonstrating the need for turning off the chirp at central time. 
			The parameters are: $\Omega_{3(0)} = 5.0[\omega_{21}]$, $\tau_0 = 10[\omega^{-1}_{21}]$, $\Delta = 1.0[\omega_{21}]$ and $\alpha_s'/\tau_0^2 = -7.5$. 
		}\label{populations}
	\end{figure}
	
	To demonstrate the selective excitation of molecules using C-CARS, the time evolution of populations $\rho_{11}$ and $\rho_{22} $ and coherence $\rho_{12}$ is presented in Fig.\,\ref{populations} for four different cases described below. The C-CARS control scheme is applied for the resonant ($\delta = 0$) and off-resonant ($\delta \neq 0$) case respectively in figures \ref{populations}(a) and \ref{populations}(b). The coherence reaches maximum at the central time in the resonant case, and is preserved till the end of dynamics owing to the zero net chirp rate attained by reversing the sign of $\alpha_p$. On the contrary, the time of the maximum coherence does not coincide with the central time in the non-resonant case, which results in a population transfer to the upper state and zero coherence. To emphasize the significance of reversing the sign of $\alpha_p$ in C-CARS control scheme we compare it with the scheme when the pump and Stokes are oppositely chirped for the whole pulse duration, $\alpha_p =-\alpha_s$. For this case the  dynamics of the system is plotted in figures \ref{populations}(c) and \ref{populations}(d) for $\delta = 0$ and $\delta = 0.1$ respectively.   In (c), even though the system reaches a perfect coherence at $t=t_c$, it drops to zero because population is  further adiabatically transferred to state $|2\rangle$. Coherence in (d) behaves similar to that of (b). 
	

	
	\begin{figure}
		\includegraphics[scale=0.6]{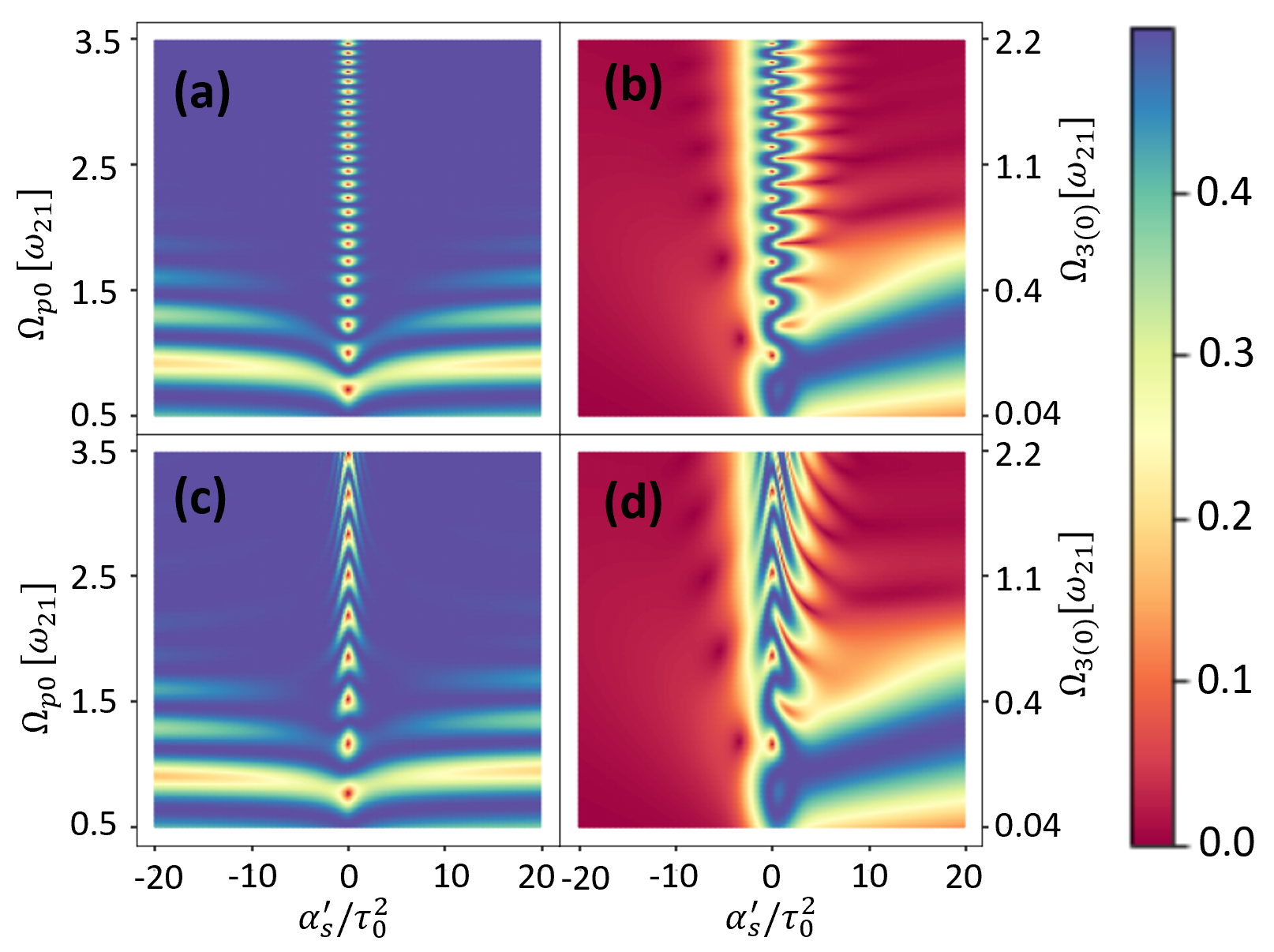}
		\caption{Vibrational coherence as a function of spectral chirp and peak Rabi frequency when C-CARS chirping scheme is used: the above figures (a and b) are plotted using super-effective two-level Hamiltonian, Eq.\,\eqref{Ham2level}, and below figures (c and d) are plotted using the exact four-level Hamiltonian, Eq.\,\eqref{Ham4level}. In figures (a) and (c) $\delta = 0$ and (b) and (d) $\delta = 0.1$. The similarity between the results of two Hamiltonians indicates the validity of adiabatic approximation which is used to derive the chirping scheme. In the case of resonance, the coherence is maximum, color blue, for most of the Rabi frequencies and spectral chirp rates, meaning that the chirping scheme is very robust against the changes in input parameters. In the absence of resonance, the coherence is zero, color red, for most values of parameters, implying that the chirping scheme is effective in selectively exciting the system. The parameters used in this figure are: $\tau_0 = 10[\omega^{-1}_{21}] $ and $\Delta_s = \Delta_{as} = 1.0[\omega_{21}]\,.$
		}\label{coherence}
	\end{figure}
	
	
	
	The validity of adiabatic approximation, which led to a derivation of the super-effective Hamiltonian, can be tested by comparing the results of the super-effective two-level system with the exact solution using the Liouville von Neumann equation for the four-level system. To this end, the condition for chirping of the probe pulse  $\alpha_{pr} = \alpha_s - \alpha_p$ is applied to the field-interaction Hamiltonian of the four level system
	
	\begin{equation}
		\begin{split}
			&\mathbf{H}_{ex}(t) \\= & \frac{\hbar}{2}
			\begin{psmallmatrix}
				2\alpha_p(t-t_c)   & 0    &  \Omega_{p_0} (t) 
				&\Omega_{as_0}(t)\\
				0 &  2[\alpha_{s}(t-t_c)-\delta] &  \Omega_{s_0} (t)  & \Omega_{pr_0} (t) \\
				\Omega_{p_0} (t)  &  \Omega_{s_0} (t) &  -2\Delta_{s}  & 0\\
				\Omega_{as_0}(t) & \Omega_{pr_0} (t)  &  0  & 2[\alpha_p(t-t_c)-\Delta_{as}]
			\end{psmallmatrix}
		\end{split}
	\end{equation}
	
	
	Figure\,\ref{coherence} shows the contour-plot of vibrational coherence $\rho_{12}$ at the end of dynamics as a function of the peak Rabi frequency $\Omega_{3(0)}(t) $ and dimensionless spectral chirp rate $\alpha_{s}'/\tau_0^2 $. Figures\,(a) and (b) represent the $\delta = 0$ and $\delta = 0.1$ cases, respectively, of the super-effective two level system, and (c) and (d) represent the same cases obtained by the exact solution of the four-level system using the same set of parameters. In all the figures, the one-photon detuning is $\Delta_s = \Delta_{as} = \Delta = 1.0$. Around the region where $\alpha_s^{'}/\tau_0^2 = 0$,  adiabatic passage breaks down and the approximate solution disagrees with the exact solution. As the magnitude of spectral chirp rate increases, the adiabatic approximation coincides with the exact one because the Landau–
	Zener parameter is well in the adiabatic range, $\Omega_{3(0)}^2/|\alpha_p|\gg 1$. In the resonant cases (a) and (c), the coherence is at the maximum, (blue color), for the most part, indicating the robustness of C-CARS chirping scheme in preparing the system in a coherent superposition. In the off-resonant case, zero coherence, (red color), is observed for the most part, which is in a stark contrast with the resonant case, revealing the selective nature of coherent excitation using the C-CARS control  scheme.  
	
	\begin{figure}
		\includegraphics[scale=0.55]{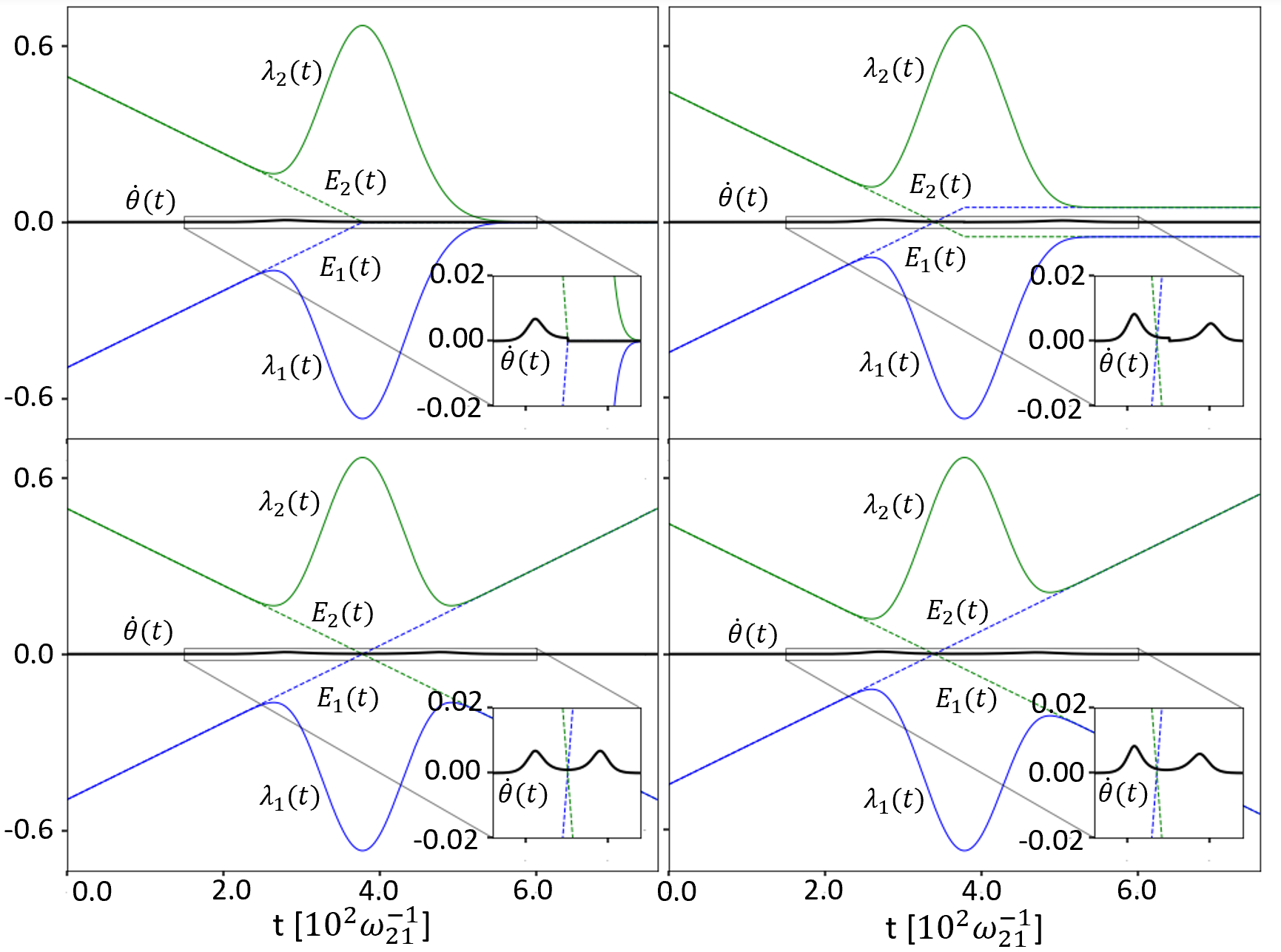}
		\caption{The evolution of the bare state and the dressed state energies and the non-adiabatic parameter: $E_1(t)$ and $E_2(t)$ (dashed lines) are the bare state energies and $\lambda_1(t)$ and $\lambda_2(t)$ (solid lines) are the dressed state energies. Figures (a) and (b) are the resonant and off-resonant cases respectively when C-CARS scheme is used. Figures (c) and (d) show the resonant and off-resonant cases, when the pump and Stokes pulses are oppositely chirped for the whole pulse duration. In contract to all the other cases, figure (a) shows the non-adiabatic parameter $\dot{\theta}(t)$, the dark solid line, remaining at zero after the central time. The parameters used are: $\Omega_{3(0)} = 5.0[\omega_{21}]$, $\tau_0 = 10[\omega^{-1}_{21}]$, $\Delta = 1.0[\omega_{21}]$ and $\alpha_s'/\tau_0^2 = -7.5$.  
		}\label{dressed_energies}
	\end{figure}
	
	When the electromagnetic field interacts with any quantum system, the eigenstates undergo a rearrangement resulting in a set of quantum states that are said to be `dressed' by the light. These states are called dressed states and the initial states that were `untouched' by the light are called bare states \cite{Berman_dressed_1982}. The robustness of the C-CARS control  scheme stems from the adiabatic nature of the light-matter interaction,  which can be demonstrated by analyzing the evolution of dressed state energies in the super-effective two-level system. To this end, the density matrix $\boldsymbol{\rho}(t) $ is transformed to a dressed density matrix using the transformation $\boldsymbol{\rho}_d(t) = \mathbf{T}(t)\boldsymbol{\rho}(t) \mathbf{T}^{\dagger}(t), $ where $\mathbf{T}(t) $ is an orthogonal matrix given by:
	\begin{equation}\label{T}
			\mathbf{T}(t) = 
			\begin{pmatrix}
				\cos\theta(t)	&	-\sin\theta(t)	\\
				\sin\theta(t)	&	\cos\theta(t) 	\\
			\end{pmatrix}\,.
	\end{equation}
	
	The Liouville-von Neumann equation for the dressed density matrix is derived from the bare state density matrix equation and reads
	\begin{eqnarray}
		i\hbar \dv{t}\boldsymbol{\rho}(t) &=& i\hbar \dv{t}\left(\mathbf{T}^{\dagger}(t)\boldsymbol{\rho}_d(t) \mathbf{T}(t)\right)\nonumber\\
		&=& \left[\mathbf{H}_{se}(t),\, \mathbf{T}^{\dagger}(t)\boldsymbol{\rho}_d(t) \mathbf{T}(t)\right]\, .
	\end{eqnarray}
	After expanding this equation and using
	$\dot{\mathbf{T}}(t)\mathbf{T^{\dagger}}(t) = -\mathbf{T}(t)\dot{\mathbf{T}}^{\dagger}(t) \, ,$ we arrive at 
	\begin{eqnarray}
		i\hbar \dot{\boldsymbol{\rho}}_d(t) &=& \left[\mathbf{T}(t)\mathbf{H}_{se}(t)\mathbf{T^ \dagger}(t)- i\hbar\mathbf{T}(t)\dot{\mathbf{T}}^{\dagger}(t),\,\boldsymbol{\rho}_d(t)\right]\nonumber\\ 
		&=& \left[\mathbf{H}_{d}(t),\, \boldsymbol{\rho}_d(t)\right]\,,
	\end{eqnarray}
	where $\mathbf{T}(t)\mathbf{H}_{se}(t)\mathbf{T^ \dagger}(t)$ is a diagonal matrix. For adiabatic passage to occur, the dressed state Hamiltonian $\mathbf{H}_d(t)$ should give the dressed state energies separation greater than  $\mathbf{T}(t)\mathbf{\dot{T}^{\dagger}}(t)$ to avoid coupling between dressed states \cite{Berman_dressed_1982, Bergmann_1989, STIRAP_Shore_2015}. The dressed state Hamiltonian is found to be
	
	\begin{widetext}
		\begin{equation}\label{dressed_Hamiltonian}
				\mathbf{H}_d(t) = \frac{\hbar}{2}
				\begin{pmatrix}
					-\sqrt{(-\delta+\alpha_{pr}(t-t_{c}))^2+(2\Omega_{3}(t))^2}	&	-i\dot{\theta}(t)	\\
					i\dot{\theta}(t)	&	\sqrt{(-\delta+\alpha_{pr}(t-t_{c}))^2+(2\Omega_{3}(t))^2}	\\
				\end{pmatrix}\,,
			\end{equation}
		\end{widetext}
		where the non-adiabatic parameter $\dot{\theta}(t)$, which comes from the matrix $\mathbf{T}(t)\mathbf{\dot{T}^{\dagger}}(t)$, is given by
		\begin{align}
			\dot\theta(t) &=  \frac{(-\delta+\alpha_{pr}(t-t_{c}))\dot\Omega_3(t) - 2\Omega_3(t)\alpha_{pr}}{(-\delta+\alpha_{pr}(t-t_{c}))^2 + 4\Omega_3(t)^2}\,.
		\end{align}
		
		\begin{figure}
			\includegraphics[scale=0.33]{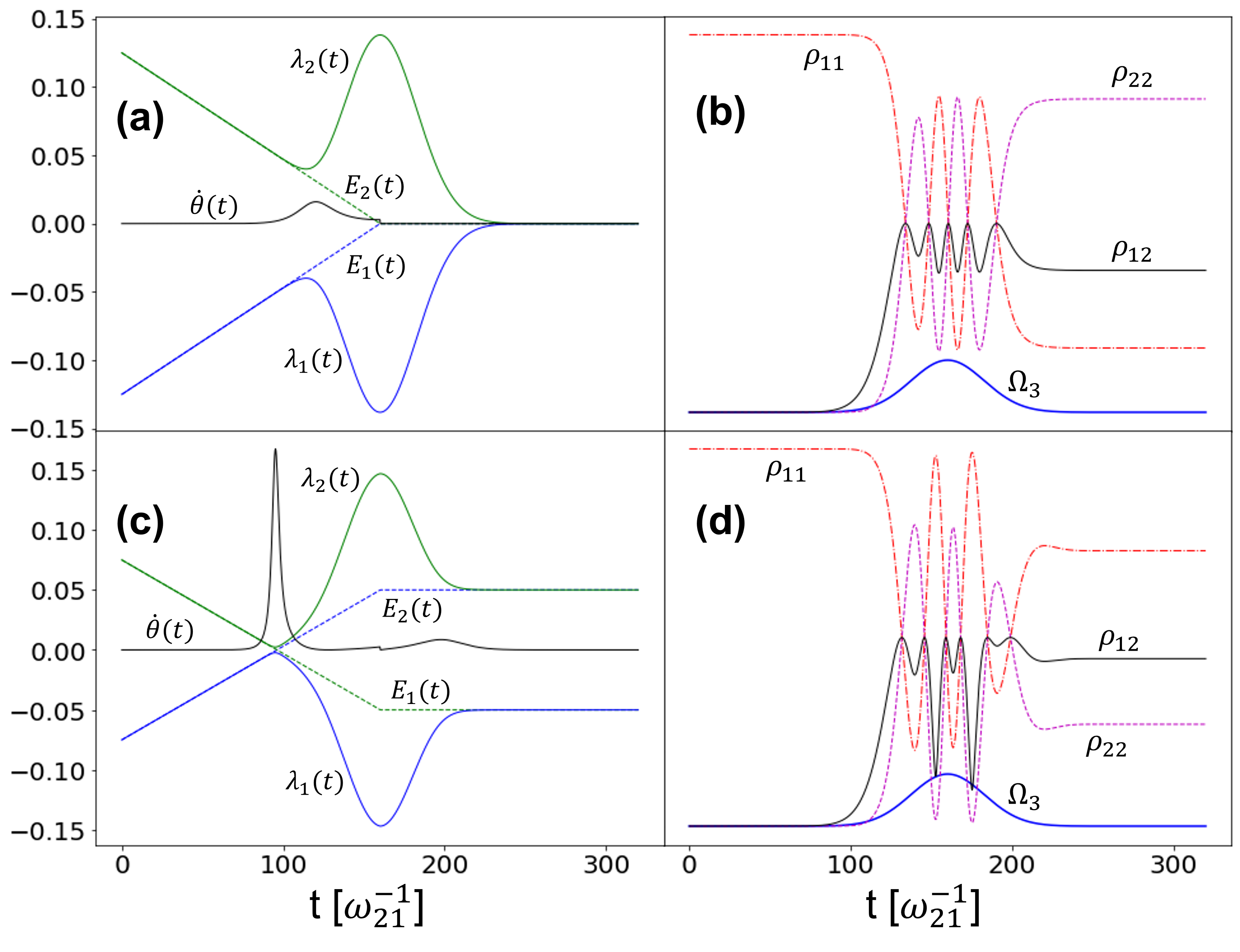}
			\caption{Evolution of energies, populations and coherence for for field parameters for a weak Rabi frequency. Here the parameters are: $\Omega_{3(0)} = 0.18$, $\tau_0 = 25$ and $\alpha_s'/\tau_0^2 = -0.8$. Figure (a) show the dressed and bare state energies and (b) show the population dynamics in the case of resonance ($\delta =0$). Figures (c) and (d) show the same plots in the absence of resonance ($\delta =0.1 $).
			}\label{populations-2}
		\end{figure}
		
		Analyzing the equation for $\dot{\theta}(t)$ reveals the selective nature of adiabatic passage in the case of resonance. In the resonant case, the C-CARS chirping scheme ensures that the process be adiabatic in the second half of the pulse by keeping the  non-adiabatic coupling parameter $\dot{\theta}(t)$ at zero during this time period. But adiabaticity is not guaranteed in the off-resonant case due to the non-zero factor $\delta$ in the equation. This is demonstrated in Fig.\, \ref{dressed_energies}, where the bare state energies are given by $E_1(t) = {H}_{se_{11}}(t)$ and  $E_2(t) = {H}_{se_{22}}(t)$  and the dressed state energies are given by  $\lambda_1(t) = {H}_{d_{11}}(t)$ and $\lambda_2(t) = {H}_{d_{22}}(t)$. The figures (a) and (b) represent resonant ($\delta =0$) and off-resonant ($\delta \neq 0$) cases when C-CARS control scheme is used. Clearly, the $\dot{\theta}(t)$, dark sold line, has non-zero values in the second half when the system is detuned. The perfectly adiabatic nature of interaction in Fig.\,\ref{dressed_energies}(a) corresponds to the maximum coherence in Fig.\,\ref{populations}(a) and the non-adiabatic nature in Fig.\,\ref{dressed_energies}(b) corresponds to the population inversion in Fig.\,\ref{populations}(b). The parameters used in Fig.\,\ref{dressed_energies} are the same as that used in Fig.\,\ref{populations}.  In figures \ref{dressed_energies}(c) and \ref{dressed_energies}(d), the same quantities are plotted for $\delta =0$ and $\delta \neq 0$ respectively for the scheme when the pump and Stokes are chirped oppositely for the whole pulse duration. 
		The process is perfectly adiabatic only in the second half of (a) since a smooth realization of $\dot{\theta}(t) = 0$ was made possible owing to the developed C-CARS control scheme. The dynamics is much different and the selective excitation does not happen when the effective Rabi frequency, $\Omega_3(t)$, is not strong enough as shown in Fig.\,\ref{populations-2} where $\Omega_3(0) = 0.18$, $\alpha_s'/\tau_0^2 = -0.8$ and $\tau_0 = 25[\omega^{-1}_{21}]$. In both (a), $\delta=0$ and (b), $\delta=0.1 $ cases, the non-adiabatic coupling parameter is much greater compared to Fig.\,\ref{dressed_energies}. In the off-resonant case (d), the coherence oscillates much stronger compared to the resonant case (c) showing the absence of adiabatic passage.
		
\begin{figure}
			\includegraphics[scale=1.0]{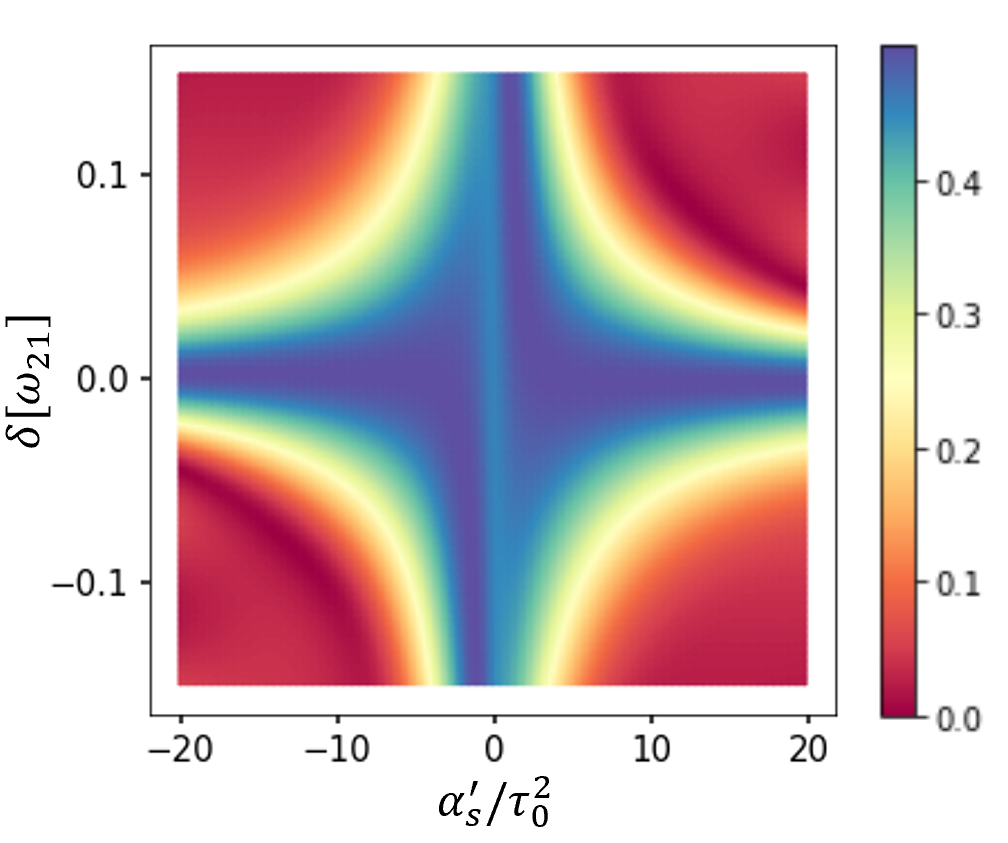}
			\caption{Coherence as a function of two-photon detuning and dimensionless spectral chirp rate. As the chirp rate increases, the system becomes selective to lesser values of $\delta$. For the values of spectral chirp rate close to zero, $\alpha_s^{\prime}/\tau_0^2 \approx 0 $, a vertical line is present as the selectivity is degraded where effectively there is no chirping.
				Flipping the signs of detuning and chirp rates results in flipping of energy levels in the two-level Hamiltonian, creating a symmetry for the coherence values of ($\alpha_s^{\prime}/\tau_0^2$, $\delta$) and ($-\alpha_s^{\prime}/\tau_0^2$, $-\delta$). The parameters used in the figure are: $\Omega_{3(0)}=1.6[\omega_{21}]$ and $\tau_0 = 4.66[\omega^{-1}_{21}]$. 
			}\label{delta_vs_chirp}
		\end{figure}

		
		To investigate how the value of two-photon detuning and spectral chirp rate are related to the selectivity in C-CARS method, the end-of-pulse vibrational coherence is plotted as a function of $\delta $ and $\alpha_s'/\tau_0^2 $ in Fig.\,\ref{delta_vs_chirp}. At the two-photon resonance,  coherence is at the maximum for all the values of chirp rate. If the detuning is large, a small chirp rate can selectively excite the system while for smaller values of detuning, the chirp rate needs to be increased in order to suppress excitation of close  vibrational frequencies and the non-resonant background. This provides a way to control the selectivity by adjusting the values of chirp rate. The plot is symmetric across the diagonal lines as flipping the signs of both $\alpha_{s}$ and $\delta$ will not change the dynamics; it would only result in switching of the diagonal elements in Hamiltonian \eqref{Ham2level}. The plot is also nearly symmetric across the $\delta = 0$ line, indicating that the selectivity holds for both red and blue detunings. When the chirp rate is close to zero, the pulses are transform-limited, implying that the maximum coherence and selectivity provided by the chirping scheme is absent in this region. This explains the vertical line present at the 0 of abscissa.
		
		In summary, we presented a control scheme to prepare the ground electronic-vibrational states in the four-level system of CARS in a maximally coherent superposition. We derived the Hamiltonian for a `super-effective' two-level system employing the adiabatic approximation. This two-level Hamiltonian is used to derive the conditions for adiabatic passage necessary for the implementation of the selective excitation of spectrally close vibrations. The amplitudes of the Stokes and probe pulses have to be equal and should be $\sqrt{2} $ times that of the pump pulse. The pump pulse should be chirped at the same rate as the Stokes pulse before the central time and at opposite rate after that. The probe pulse has to be chirped at a rate equal the difference between the chirp rates of the Stokes and the pump pulses for the whole pulse duration. The solutions of Liouville-von Neumann equation show that vibrational coherence is preserved until the end of dynamics in the resonant case due to the adiabatic nature of the interaction. At two-photon resonance, vibrational coherence is maximum, 0.5, for a wide range of field parameters revealing the robustness of the method. Conversely, coherence is almost zero in the off-resonant case for most of the peak Rabi frequency values and the chirp parameters. A comparison of coherence in the four-level and the two-level systems reveals that the adiabatic approximation is valid except when the chirp rate is almost equal to zero. A dressed-state analysis reveals the details of the mechanism of adiabatic passage using C-CARS control scheme. 
		
		The presented C-CARS method can find important applications in sensing and imaging of molecular species because it creates a maximally coherent superposition of vibrational states in coherent anti-Stokes Raman scattering allowing the system to emit an optimized signal suitable for detection. The robustness of this method against changes in Rabi frequencies and chirp rates is helpful in experiments. The method helps suppress the background species and excite only the desired ones; the resolution needed for this distinction can be controlled by the chirp parameter.
		
		
		\section*{Acknowledgment}
		
		S.M. and J.Ch. acknowledge support from the Office of Naval Research under awards N00014-20-1-2086 and N00014-22-1-2374. S.M. acknowledges the Helmholtz Institute Mainz Visitor Program. The work of D.B. was supported in part by the European Commission’s Horizon Europe Framework Program under the Research and Innovation Action MUQUABIS GA no. 101070546.

		
		{}
		

\begin{thebibliography}{}
			
			\bibitem{XieCRSMicroscopy}Cheng J X and Xie X S 2013 {\em Coherent Raman Scattering Microscopy,} Taylor \& Francis Group, LLC
			
			\bibitem{Xie_CARS_2008}Evans, C L and Xie, X S 2008 Coherent Anti-Stokes Raman Scattering Microscopy: Chemical Imaging for Biology and Medicine. \textit{Annual Review of Analytical Chemistry} \textbf{1} 883–909.
			
			\bibitem{Xie_CARS_2004} Cheng J X and Xie X S 2004 Coherent Anti-Stokes Raman Scattering Microscopy: Instrumentation, Theory, and Applications. \textit{J. Phys. Chem. B} \textbf{108} 827-840
			
			\bibitem{Quantitative_CARS_2011}  Day J P, Domke K F, Rago G, Kano H, Hamaguchi H-o, Vartiainen E M and Bonn M 2011 Quantitative coherent anti-stokes raman scattering (CARS) microscopy {\em J. Phys. Chem. B} {\bf 115} 7713–25
			
			\bibitem{Polymer_2022}Xu S, Camp C H, Lee Y J 2022 Coherent anti-Stokes Raman scattering microscopy for polymers. \textit{J. Polym. Sci} \textbf{60} 1244-65.
			
			\bibitem{Xie_1999}Zumbusch A, Holtom G R and Xie X S 1999 Three-Dimensional Vibrational Imaging by Coherent Anti-Stokes Raman Scattering \textit{Phys. Rev. Lett.} \textbf{82}, 4142–4145
			
			\bibitem{Xie_imaging_2005}Evans C L, Potma E O, Puoris'haag M, Côté D, Lin C P and Xie X S 2005 Chemical imaging of tissue \textit{in vivo} with video-rate coherent anti-stokes raman scattering microscopy \textit{Proceedings of the National Academy of Sciences} \textbf{102} 16807–12			  
			
			\bibitem{Xie_laser_2002} Cheng J-X, Jia Y K, Zheng G and Xie X S 2002 Laser-scanning coherent anti-stokes raman scattering microscopy and applications to Cell Biology \textit{Biophysical Journal} \textbf{83} 502–9			
			
			\bibitem{Xie_imaging_2001}
			
			Volkmer A, Cheng J-X and Sunney Xie X 2001 Vibrational imaging with high sensitivity via epidetected coherent Anti-Stokes Raman scattering microscopy \textit{Phys. Rev. Lett.} \textbf{87} 023901
			
			\bibitem{Potma_hydro_2001} Potma E O, de Boeij W P, van Haastert P J and Wiersma D A 2001 Real-time visualization of intracellular hydrodynamics in single living cells \textit{Proceedings of the National Academy of Sciences} \textbf{98} 1577–82 
			
			\bibitem{Potma_imaging_2006} Nan X, Potma E O and Xie X S 2006 Nonperturbative chemical imaging of organelle transport in living cells with coherent anti-stokes raman scattering microscopy {\em Biophysical Journal} {\bf 91} 728–35 
			
			\bibitem{Potma_lipid_2003}  Potma E O and Xie X S 2003 Detection of single lipid bilayers with coherent anti-stokes raman scattering (CARS) microscopy {\em Journal of Raman Spectroscopy} {bf 34} 642–50
			
			\bibitem{Potma_imaging_2004} Potma E O, Xie X S, Muntean L, Preusser J, Jones D, Ye J, Leone S R, Hinsberg W D and Schade W 2003 Chemical imaging of photoresists with coherent anti-stokes raman scattering (CARS) microscopy {\em J. Phys. Chem. B} {\bf 108} 1296–301 
			
			\bibitem{Potma_imaging_2010}  Balu M, Liu G, Chen Z, Tromberg B J and Potma E O 2010 Fiber delivered probe for efficient cars imaging of tissues {\em Optics Express} {\bf 18} 2380
			
			\bibitem{Saykally_imaging_2002} Schaller R D, Ziegelbauer J, Lee L F, Haber L H and Saykally R J 2002 Chemically selective imaging of subcellular structure in human hepatocytes with coherent Anti-Stokes Raman scattering (CARS) near-field scanning optical microscopy (NSOM) {\em J. Phys. Chem. B} {\bf 106} 8489–92
			
			\bibitem{nature_imaging_2014}  Camp Jr C H, Lee Y J, Heddleston J M, Hartshorn C M, Walker A R, Rich J N, Lathia J D and Cicerone M T 2014 High-speed coherent Raman fingerprint imaging of biological tissues {\em Nature Photonics} {\bf 8} 627–34
			
			\bibitem{Dantus_standoff_2008}  Li H, Harris D A, Xu B, Wrzesinski P J, Lozovoy V V and Dantus M 2008 Coherent mode-selective Raman excitation towards standoff detection {\em Optics Express} {\bf 16} 5499
			
			\bibitem{Dantus_standoff_2013}  Bremer M T and Dantus M 2013 Standoff explosives trace detection and imaging by selective stimulated Raman scattering {\em Applied Physics Letters} {\bf 103} 061119
			
			\bibitem{Sokolov_detection_2005}  Petrov G I, Yakovlev V V, Sokolov A V and Scully M O 2005 Detection of bacillus subtilis spores in water by means of broadband coherent anti-stokes raman spectroscopy {\em Optics Express} {\bf 13} 9537
			
			\bibitem{scully-cars}  Ooi C H, Beadie G, Kattawar G W, Reintjes J F, Rostovtsev Y, Zubairy M S and Scully M O 2005 Theory of femtosecond coherent anti-stokes raman backscattering enhanced by quantum coherence for standoff detection of bacterial spores {\em Phys. Rev. A} {\bf 72} 023807
			
			\bibitem{Gord_CARS_2010}  Roy S, Gord J R and Patnaik A K 2010 Recent advances in coherent anti-stokes raman scattering spectroscopy: Fundamental developments and applications in reacting flows {\em Progress in Energy and Combustion Science} {\bf 36} 280–306
			
			\bibitem{Gord_comparison_2017}  Richardson D R, Stauffer H U, Roy S and Gord J R 2017 Comparison of chirped-probe-pulse and hybrid femtosecond/picosecond coherent anti-stokes raman scattering for combustion thermometry {\em Applied Optics} {\bf 56}
			
			\bibitem{cancer1} Petrov G I, Arora R and Yakovlev V V 2021 Coherent anti-stokes raman scattering imaging of microcalcifications associated with breast cancer {\em The Analyst} {\bf 146} 1253–9
			
			\bibitem{cancer2} Aljakouch K, Hilal Z, Daho I, Schuler M, Krauß S D, Yosef H K, Dierks J, Mosig A, Gerwert K and El-Mashtoly S F 2019 Fast and noninvasive diagnosis of cervical cancer by coherent anti-stokes raman scattering {\em Analytical Chemistry} {\bf 91} 13900–6
			
			\bibitem{cancer3}  Legesse F B, Medyukhina A, Heuke S and Popp J 2015 Texture analysis and classification in coherent anti-stokes raman scattering (CARS) microscopy images for automated detection of skin cancer {\em Computerized Medical Imaging and Graphics} {\bf 43} 36–43
			
			\bibitem{COVID} Tabish T A, Narayan R J and Edirisinghe M 2020 Rapid and label-free detection of covid-19 using coherent anti-stokes raman scattering microscopy {\em MRS Communications} {\bf 10} 566–72
			
			\bibitem{Potma_real-time_2014} Yampolsky S, Fishman D A, Dey S, Hulkko E, Banik M, Potma E O and Apkarian V A 2014 Seeing a single molecule vibrate through time-resolved coherent anti-stokes raman scattering {\em Nature Photonics} {\bf 8} 650–6
			
			\bibitem{nature_symmetry_2016} Cleff C, Gasecka A, Ferrand P, Rigneault H, Brasselet S and Duboisset J 2016 Direct imaging of molecular symmetry by coherent anti-stokes Raman scattering {\em Nature Communications} {\bf 7} 
			
			\bibitem{nature_graphere_2019} Virga A, Ferrante C, Batignani G, De Fazio D, Nunn A D, Ferrari A C, Cerullo G and Scopigno T 2019 Coherent anti-stokes raman spectroscopy of single and multi-layer graphene {\em Nature Communications} {\bf 10}
			
			\bibitem{CARS_multiplex_2014} Bohlin A and Kliewer C J 2014 Two-beam ultrabroadband coherent anti-stokes raman spectroscopy for high resolution gas-phase multiplex imaging {\em Applied Physics Letters} {\bf 104} 031107		
			
			\bibitem{flame_wall_2015} Bohlin A, Mann M, Patterson B D, Dreizler A and Kliewer C J 2015 Development of two-beam femtosecond/picosecond one-dimensional rotational coherent anti-stokes raman spectroscopy: Time-resolved probing of flame wall interactions {\em Proceedings of the Combustion Institute} {\bf 35} 3723–30
			
			\bibitem{review_skeletal_2016} Moura C C, Tare R S, Oreffo R O and Mahajan S 2016 Raman spectroscopy and coherent anti-stokes raman scattering imaging: Prospective tools for monitoring skeletal cells and skeletal regeneration {\em Journal of The Royal Society Interface} {\bf 13} 20160182 
			
			\bibitem{Zhetlikov_2000} Zheltikov A M 2000 Coherent anti-stokes raman scattering: From proof-of-the-principle experiments to femtosecond cars and higher order wave-mixing generalizations {\em Journal of Raman Spectroscopy} {\bf 31} 653–67
			
			\bibitem{Levis_BOXCARS_2007} Romanov D, Filin A, Compton R and Levis R 2007 Phase matching in femtosecond boxcars {\em Optics Letters} {\bf 32} 3161
			
			\bibitem{Scully_coherent_vs_incoherent_2007} Pestov D, Ariunbold G O, Murawski R K, Sautenkov V A, Sokolov A V, and Scully M O 2007 Coherent versus incoherent Raman scattering: molecular coherence excitation and measurement, {\em Opt. Lett.} {\bf 32}, 1725
			
			\bibitem{Xie_SRS_2008} Freudiger C W, Min W, Saar B G, Lu S, Holtom G R, He C, Tsai J C, Kang J X and Xie X S 2008 Label-free biomedical imaging with high sensitivity by stimulated Raman scattering microscopy {\em Science} {\bf 322} 1857–61
			
			\bibitem{CARS_tutorial_2012}Krafft C, Dietzek B, Schmitt M and Popp, J 2012 Raman and Coherent Anti-Stokes Raman Scattering Microspectroscopy for Biomedical Applications. \textit{Journal of Biomedical Optics}, \textbf{17} 040801.
			
			\bibitem{Potma_background-free_2006} Potma E O, Evans C L and Xie X S 2006 Heterodyne coherent anti-stokes raman scattering (CARS) imaging {\em Optics Letters} {\bf 31} 241
			
			\bibitem{Potma_background-free_2004} Evans C L, Potma E O and Xie X S 2004 Coherent anti-Stokes Raman scattering spectral interferometry: determination of the real and imaginary components of nonlinear susceptibility $\chi^{(3)} $ for vibrational microscopy {\em Optics Letters} {\bf 29} 2923
			
			\bibitem{background-free_2003} Oron D, Dudovich N and Silberberg Y 2003 Femtosecond phase-and-polarization control for background-free coherent anti-stokes raman spectroscopy {\em Physical Review Letters} {\bf 90} 213902
			
			\bibitem{background-free_2008} Gachet D, Billard F and Rigneault H 2008 Focused field symmetries for background-free coherent anti-stokes Raman spectroscopy {\em Phys. Rev. A} {\bf 77} 061802
			
			\bibitem{background-free_2010} Selm R, Winterhalder M, Zumbusch A, Krauss G, Hanke T, Sell A and Leitenstorfer A 2010 Ultrabroadband background-free coherent anti-stokes raman scattering microscopy based on a compact ER:Fiber Laser System {\em Optics Letters} {\bf 35} 3282
			
			\bibitem{background-free_2010-2}  Konorov S O, Blades M W and Turner R F 2010 Lorentzian amplitude and phase pulse shaping for nonresonant background suppression and enhanced spectral resolution in coherent Anti-Stokes Raman scattering spectroscopy and microscopy {\em Applied Spectroscopy} {\bf 64} 767–74
			
			\bibitem{nature_high-resolution_2018}  Lombardini A, Mytskaniuk V, Sivankutty S, Andresen E R, Chen X, Wenger J, Fabert M, Joly N, Louradour F, Kudlinski A and Rigneault H 2018 High-resolution multimodal flexible coherent Raman endoscope {\em Light, Science \& Applications} {\bf 7}
			
			\bibitem{chirped-probe-pulse_2011}  Richardson D R, Lucht R P, Kulatilaka W D, Roy S and Gord J R 2011 Theoretical modeling of single-laser-shot, chirped-probe-pulse femtosecond coherent anti-stokes raman scattering thermometry {\em Appl. Phys. B} {\bf 104} 699–714
			
			\bibitem{Pestov_Optim_2007}  Pestov D, Murawski R K, Ariunbold G O, Wang X, Zhi M, Sokolov A V, Sautenkov V A, Rostovtsev Y V, Dogariu A, Huang Y and Scully M O 2007 Optimizing the laser-pulse configuration for coherent Raman spectroscopy {\em Science} {\bf 316} 265–8
			
			\bibitem{Kumar_background-free_2011}  Kumar V, Osellame R, Ramponi R, Cerullo G and Marangoni M 2011 Background-free broadband CARS spectroscopy from a 1-MHz ytterbium laser {\em Optics Express} {\bf 19} 15143
			
			\bibitem{Ch21}  Chathanathil J, Liu G and Malinovskaya S A 2021 Semiclassical control theory of coherent anti-stokes raman scattering maximizing vibrational coherence for remote detection {\em Phys. Rev. A} {\bf 104} 043701
			
			\bibitem{Saykally_chirped_2006}  Knutsen K P, Messer B M, Onorato R M and Saykally R J 2006 Chirped coherent anti-stokes raman scattering for high spectral resolution spectroscopy and chemically selective imaging {\em J. Phys. Chem. B} {\bf 110} 5854–64
			
			\bibitem{Saykally_chirped_2007}  Onorato R M, Muraki N, Knutsen K P and Saykally R J 2007 Chirped coherent anti-stokes raman scattering as a high-spectral- and spatial-resolution microscopy {\em Optics Letters} {\bf 32} 2858
			
			\bibitem{Pandya20}  Pandya N, Liu G, Narducci F A, Chathanathil J and Malinovskaya S 2020 Creation of the maximum coherence via adiabatic passage in the four-wave mixing process of coherent anti-stokes raman scattering {\em Chem. Phys. Lett.} {\bf 738} 136763
			
			\bibitem{Malinovsky_2009}  Malinovsky V S 2009 Creating maximum coherence using chirped pulses without adiabatic elimination of excited states {\em Journal of Raman Spectroscopy} {\bf 40} 817–21
			
			\bibitem{Ma07} Malinovskaya S A 2007 Chirped Pulse Control Methods for Imaging of Biological Structure and Dynamics {\em Int. J. Quant. Chem.} {\bf 107} 3151
			
			\bibitem{Ma2006} Malinovskaya S A 2006 Mode-selective excitation using ultrafast chirped laser pulses {\em Phys. Rev. A} {\bf 73}
			
			\bibitem{Patel-2011} Patel V and Malinovskaya S A 2011 Nonadiabatic effects induced by the coupling between vibrational modes via Raman fields {\em Phys. Rev. A} {\bf 83}
			
			\bibitem{Ma01} Malinovskaya S A and Malinovsky V S 2007 Chirped-pulse adiabatic control in co- herent anti-Stokes Raman scattering for imaging	of biological structure and dynamics {\em Optics Letters} {\bf 32} 707
			
			\bibitem{Berman_dressed_1982} Berman P R and Salomaa R 1982 Comparison between dressed-atom and bare-atom pictures in laser spectroscopy {\em Phys. Rev. A} {\bf 25} 2667–92
			
			\bibitem{Bergmann_1989}  Kuklinski J R, Gaubatz U, Hioe F T and Bergmann K 1989 Adiabatic population transfer in a three-level system driven by delayed laser pulses {\em Phys. Rev. A} {\bf 40} 6741–4
			
			\bibitem{STIRAP_Shore_2015}  Bergmann K, Vitanov N V and Shore B W 2015 Perspective: Stimulated raman adiabatic passage: The status after 25 Years {\em The Journal of Chemical Physics} {\bf 142} 170901
			
			
			
			
			
			
		\end{thebibliography}
	\end{document}